\documentclass[lettersize,journal]{IEEEtran}
\usepackage{amsmath,amsfonts}
\usepackage{algorithmic}
\usepackage{algorithm}
\usepackage{array}
\usepackage[caption=false,font=normalsize,labelfont=sf,textfont=sf]{subfig}
\usepackage{textcomp}
\usepackage{stfloats}
\usepackage{url}
\usepackage{verbatim}
\usepackage{graphicx}
\usepackage{color}
\usepackage{cite}
\begin{document}

\title{A Human-Machine Joint Learning Framework to Boost Endogenous BCI Training}

\author{Hanwen Wang, Yu Qi*,~\IEEEmembership{Member,~IEEE,} Lin Yao, Yueming Wang,~\IEEEmembership{Member,~IEEE,}\\ Dario Farina,~\IEEEmembership{Fellow,~IEEE,} Gang Pan*,~\IEEEmembership{Member,~IEEE,}
\thanks{Hanwen Wang is with the College of Computer Science and Technology, Zhejiang University, Hangzhou, China.}
\thanks{Yu Qi is with the Affiliated Mental Health Center \& Hangzhou Seventh People’s Hospital, and the MOE Frontier Science Center for Brain Science and Brain-machine Integration, Zhejiang University School of Medicine, the State Key Lab of Brain-Machine Intelligence, Hangzhou, China.}
\thanks{Lin Yao is with the Department of Neurobiology, Affiliated Mental Health Center \& Hangzhou Seventh People's Hospital, Zhejiang University School of Medicine, MOE Frontiers Science Center for Brain and Brain-Machine Integration, Zhejiang University, Department of Biomedical Engineering, Zhejiang University, and College of Computer Science and Technology, Zhejiang University, Hangzhou, China}
\thanks{Yueming Wang is with the Qiushi Academy for Advanced Studies, Zhejiang University, Hangzhou, China.}
\thanks{Dario Farina is with the Department of Bioengineering, Imperial College London, London, UK}
\thanks{Gang Pan is with the State Key Lab of Brain-Machine Intelligence, the College of Computer Science and Technology, Zhejiang University, Hangzhou, China, and the First Affiliated Hospital, Zhejiang University, Hangzhou, China.}
\thanks{* Corresponding authors: Yu Qi (qiyu@zju.edu.cn) and Gang Pan (gpan@zju.edu.cn).}}

\markboth{Journal of \LaTeX\ Class Files,~Vol.~14, No.~8, August~2021}%
{Shell \MakeLowercase{\textit{et al.}}: A Sample Article Using IEEEtran.cls for IEEE Journals}


\maketitle

\begin{abstract}
	Brain-computer interfaces (BCIs) provide a direct pathway from the brain to external devices and have demonstrated great potential for assistive and rehabilitation technologies. Endogenous BCIs based on electroencephalogram (EEG) signals, such as motor imagery (MI) BCIs, can provide some level of control. However, mastering spontaneous BCI control requires the users to generate discriminative and stable brain signal patterns by imagery, which is challenging and is usually achieved over a very long training time (weeks/months). Here, we propose a human-machine joint learning framework to boost the learning process in endogenous BCIs, by guiding the user to generate brain signals towards an optimal distribution estimated by the decoder, given the historical brain signals of the user. To this end, we firstly model the human-machine joint learning process in a uniform formulation. Then a human-machine joint learning framework is proposed: 1) for the human side, we model the learning process in a sequential trial-and-error scenario and propose a novel ``copy/new'' feedback paradigm to help shape the signal generation of the subject toward the optimal distribution; 2) for the machine side, we propose a novel adaptive learning algorithm to learn an optimal signal distribution along with the subject's learning process. Specifically, the decoder reweighs the brain signals generated by the subject to focus more on ``good'' samples to cope with the learning process of the subject.
	Online and psuedo-online BCI experiments with 18 healthy subjects demonstrated the advantages of the proposed joint learning process over co-adaptive approaches in both learning efficiency and effectiveness. 
	
	
\end{abstract}


\section{Introduction}
\label{sec:introduction}

Brain-computer interfaces (BCIs) act as an intermediary between the brain and external devices, translating brain signals to control signals \cite{shih2012brain,pan2018rapid}. BCIs employing electroencephalogram (EEG) signals have been developed successfully for many applications, including neural rehabilitation, prosthesis control and emotion recognition \cite{ang2016eeg,lazarou2018eeg,cincotti2012eeg,li2019multisource,pancholi2022source,xu2013enhanced,qi2019dynamic,gu2022frame}. 
Most existing BCI systems can be divided into endogenous and exogenous BCIs \cite{bhattacharya2000complexity,di2002cortical}. Exogenous BCIs such as the P300 and SSVEP systems, rely on brain signals evoked by external stimuli \cite{yijun2005brain,wang2021rcit}, such as visual flashes or audio events. {\color{black}Conversely, endogenous BCIs, such as based on motor imagery (MI), rely on self-regulated brain signals often induced by users’ covert attention or mental tasks \cite{hu2015comparison,gao2014visual}.} Thus, endogenous BCIs can provide more flexible and natural brain control, resulting in high potential for a wide range of applications \cite{choi2008control,doud2011continuous,jiang2015brain,qi2021learning}. In the remainder of this paper, we will only focus on endogenous BCIs.

Effective BCI control relies on the close collaboration of the brain (the subject) and the machine (the decoder)\cite{wu2013convergence}. The user should generate discriminative and sufficient brain signals as different control commands, while the decoder should identify different brain signal patterns and robustly interpret/map them into control commands. For endogenous BCIs,
the main challenge lies in the difficulty for the user in generating effective and stable brain signals. To achieve robust online control, the BCI user should be able to accurately replicate at least two discriminative and stable patterns for basic control, such as going left and right. A typical endogenous BCI is based on MI, where the user is asked to imagine movements, for example of the left/right hand. However, generating MI is challenging for most users, and a typical user usually requires a long and difficult training process to learn the MI control. Commonly, training for MI lasts several weeks or even months \cite{edelman2019noninvasive,hwang2009neurofeedback}. Furthermore, there are at least 15\% to 30\% of subjects across the population who cannot control BCI systems through learning, a case called BCI illiteracy \cite{vidaurre2010towards}. For these reasons, the facilitation of the learning process in endogenous BCIs is a very relevant problem. 

\begin{figure}[!t]
	\centering
	\includegraphics[width=1\columnwidth]{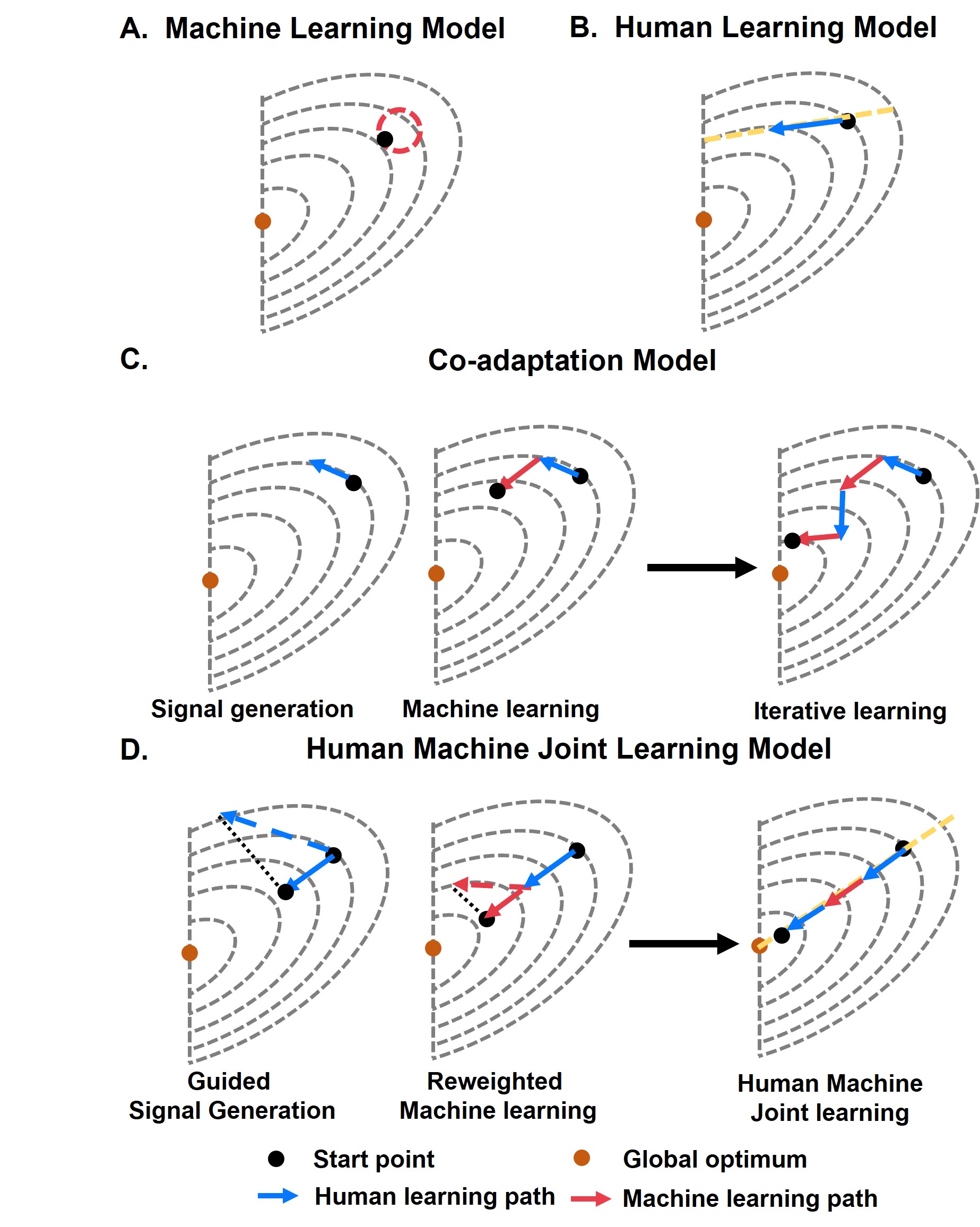} 
	\caption{Illustration of BCI training with different strategies. (A) The process of decoder learning gives a set of subject-generated data (indicated by the red circle). (B) The subject learning process with feedback from a fixed decoder. (C) The co-adaptation learning process is where the subject and the decoder learn in an alternate manner. (D) The proposed joint learning process is where the subject and the decoder share the same loss function during learning.}
	\label{different models}
\end{figure}

The learning process in BCIs includes two parts: the learning of the decoder (machine learning), and the learning of the user (human learning). Here we briefly review the existing approaches. 
\begin{itemize}
	\item
	\textbf{Machine learning.} Traditional BCI training processes start with a static subject learning process where the subject tries to generate brain signals without feedback. Then a decoder is trained with the initial data to learn discriminative features from different brain signals (as shown in Fig. \ref{different models}A). The discrimination of different brain signal patterns can be regarded as a typical classification problem. Thus, linear and nonlinear classifiers have been proposed for the decoding problem. 
	{\color{black} For linear methods, decoding in motor imagery (MI) can be separated into feature extraction and classification. Common Spatial Pattern (CSP) is a widely-used linear approach for extracting discriminative features from different brain signal patterns. Filter Bank Common Spatial Pattern (FBCSP) \cite{ang2008filter} extends CSP with a set of band-pass filters to enhance features, and it can also be applied in neural networks \cite{9805775}. However, traditional CSP is not suitable for multiclass problems. Joint Approximate Diagonalization (JAD)-based methods are used to improve CSP \cite{kumar2020formulating}. 
	Moreover, modifications have been made to enhance CSPs to deal with the nonstationarity in EEG signals. Fuzzy covariance matrices are deployed in DivCSP-WS to extract steady features \cite{reddy2022driver}.
	Channel selection methods could also improve stability between sessions, improving the classification accuracy \cite{sadatnejad2022riemannian}. }
	
	 {\color{black} 
	 In terms of classification methods, linear discriminant analysis (LDA) is commonly used for extracted features \cite{zhang2013z}. To enhance robustness on non-stationary EEG signals, variations of LDA such as rLDA and BLDA have been introduced \cite{friedman1989regularized,lei2009empirical,blankertz2011single}. Support Vector Machines (SVMs) with flexible kernel design are employed with feature extraction methods in some studies \cite{subasi2010eeg,bhardwaj2015classification}. }
	{\color{black}Recently, deep learning approaches including Long Short Term Memory (LSTM) \cite{wang2018lstm,zhou2018classification,tortora2020deep} and convolutional neural networks (CNNs) \cite{du2020efficient,dai2020hs,zhang2019making,bang2021spatio,hou2022gcns}, such as EEGnet \cite{lawhern2018eegnet}, have demonstrated  effectiveness in MI classification. Combining the feature extraction methods and the deep learning approach, neural structured learning (NSL) has been proposed to train deep neural networks with feature inputs and the structured signals, which enhances the robustness \cite{gupta2022performance}.}
	\item
	\textbf{Human learning.} After the initial decoder is trained, the subject can learn to control the BCI with the feedback of the decoder, where the design of the feedback plays an important part (Fig. \ref{different models}B). 
	Traditional feedbacks usually directly reflect the decoding result, using a moving cursor or a falling ball \cite{blankertz2007non,pfurtscheller2001motor}. To improve the efficiency of subject learning, new paradigms with different types of feedback have been introduced. For example, Hwang et al. proposed an intuitive feedback that uses a surface topography of EEG signal \cite{hwang2009neurofeedback}, where the real-time spectrogram is presented to enhance the MI \cite{mihara2012neurofeedback}. Wang et al. proposed a neurofeedback training paradigm, in which the performance of MI was transferred into the distance between a needle and hands on the screen \cite{wang2019bci}. 
	{\color{black} Alternatively, auditory feedbacks of different sounds have been presented to subject depending on their MI performance \cite{nijboer2008auditory}. Nevertheless, Yao et al. proposed tactile stimulation to help subjects improve the MI activation \cite{shu2017enhanced,yao2022reducing}. Also, Ono et al. adopted exoskeleton attached to subjects, making feedbacks more sensible  \cite{ono2018enhancement}.}
	\item
	\textbf{Co-adaptive learning.} 
	The aforementioned approaches mostly focus on one side of the learning, either the signal generation by the user or the algorithm training by the machine. However, co-adaptive methods have also been introduced \cite{perdikis2020brain,millan2015brain}. Instead of using a fixed decoder after the initial training process, co-adaptive approaches retrain both the decoder and the subject iteratively in sequential learning sessions to cope with each other \cite{millan2004need,perdikis2020brain} (Fig. \ref{different models}C). 
	To this end, different decoder recalibration algorithms have been proposed using machine learning approaches such as LDA \cite{vidaurre2011co}, SVM \cite{li2008self}, probabilistic graphical models \cite{llera2012adaptive} to update the decoder with newly collected data to cope with changes in brain signals during the subject learning process. By considering the interaction between the subject and the decoder, co-adaptive approaches demonstrated superior performance in BCI learning than separate learning approaches \cite{shenoy2006towards, abu2019co}. To provide a mathematical basis of co-adaptive learning processes, previous work has modeled the process with mathematical formulations. For example, in \cite{merel2015encoder}, an encoder-decoder model was proposed, in which the co-adaptation was described by the optimization of the user's control. With the assumption of two linear learning systems, a theoretical formulation was derived by the stochastic gradient descent method in \cite{muller2017mathematical}. 

\end{itemize}

In Fig. \ref{different models}, we illustrate and compare the aforementioned learning process in an intuitive way. Suppose there is a global optimal point for the human-machine BCI learning process. With the machine learning model alone (Fig. \ref{different models}A), where the human model is static, the optimal performance is highly limited by the boundary of brain signal patterns. With the human learning model alone (Fig. \ref{different models}B), where the decoder is fixed, the subject can learn with the guidance of the decoder's feedback. The performance highly relies on the effectiveness of the decoder. Since the decoder is usually trained with the initially imperfect brain signals of the subject, it can be suboptimal and can degrade the human learning process. The co-adaptation model (Fig. \ref{different models}C) considers both machine learning and human learning, where the decoder is iteratively updated along with the human learning process. Specifically, it contains an iterative process where the machine and the human learn alternately, which improves the theoretical overall performance. However, it assumes that the subject can effectively learn at every human learning stage, which can be difficult, especially at the beginning rounds of the training process. 

This study proposes a novel BCI learning scheme for efficient and effective BCI training.
As shown in Fig. \ref{different models}D, we aim to construct a joint learning model between humans and machines, where human and machine processes can be optimized simultaneously. 
To this end, we propose a novel human-machine joint learning framework for effective BCI training. 
Specifically, we assume the human learning behavior in sequential training is a trial-and-error process, and let the decoder determine whether a brain signal is ``good'' or ``poor'' by its discrimination ability according to the distribution of the feature space.
The contribution of this study is summarized in three folds.
\begin{enumerate}
	\item
	We formally formulate the human-machine joint learning model and propose a human-machine joint loss function, where the subject is encouraged to generate more discriminative brain signals, and the decoder optimizes the classifier to separate the different brain signal modes. 
	\item
	From the human side, we formulate human behavior in a sequential trial-and-error learning process. A novel paradigm is proposed to guide the subject to optimize brain signals. Specifically, if a ``good'' signal is generated, the system encourages the subject to ``copy'' the state; otherwise, if a ``poor'' signal is generated, the subject is encouraged to change the brain signal. 
	\item
	From the machine side, we propose a novel adaptive learning algorithm to learn an optimal signal distribution along with the subject's learning process. Specifically, the decoder reweighs the brain signals generated by the subject to focus more on ``good'' samples to cope with the learning process of the subject.
\end{enumerate}

Online BCI experiments with 18 healthy subjects demonstrated that the joint learning framework can efficiently guide the subject learning discriminate brain signals for effective BCI control. Compared with traditional co-adaptive approaches, our method significantly improved the average control accuracy from 69.1\% to 74.5\%.

\section{The Human-machine joint learning framework}
We first formulate the BCI learning process in a uniform model and propose a human-machine joint loss function for learning. Then we propose the human learning model and propose a loss function for the human learning process. We then introduce the machine learning process where the decoder learns to select optimal brain signals that the subject generated to guide and cope with the learning process of the subject. Finally, we propose a novel BCI learning paradigm and framework for human-machine joint learning. 

\subsection{Modeling of the human-machine joint learning}

Considering BCI learning as a human-machine learning problem, the subject and the decoder share the same goal of accurate intention decoding from brain signals. To achieve this goal, the subject tries to generate more discriminative brain signals, and the decoder optimizes the classifier to best separate different brain signal patterns. 
To this end, the human learning model can be described as:
\begin{equation}
	x=g(\theta_{H},y),
	\label{humanfunction}
\end{equation}
where $x$ stands for the signal generated according to the given label $y$. And the parameter of generation and the function are denoted as $\theta_{H}$ and $g(.)$ respectively.
For the machine part, the model is defined by:
\begin{equation}
	y'=h(\theta_{M},x),
	\label{machinefunction}
\end{equation}
in which the predict label $y'$ is decoded by function $h(.)$ and parameter $\theta_{M}$.

Thus, the human-machine mutual goal of BCI learning is to minimize the distance between the true label $y$ and the label estimated by the decoder given brain signals $y'$:
\begin{equation}
	\begin{split}
		L_{H-M} &= ||y'-y||.
	\end{split}
	\label{machineloss}
\end{equation}
Then the joint optimization of humans and machines can be achieved by minimizing the joint loss function $L_{H-M}$:
\begin{equation}
	\begin{split}
		\mathop{\arg\min_{\theta_{H},\theta_{M}}}L_{H-M}
		& =\mathop{\arg\min_{\theta_{H},\theta_{M}}} ||h(\theta_{M},g(\theta_{H},y))-y||.
	\end{split}
	\label{machineloss}
\end{equation}

\begin{figure}[!t]
	\centering
	\includegraphics[width=1\columnwidth]{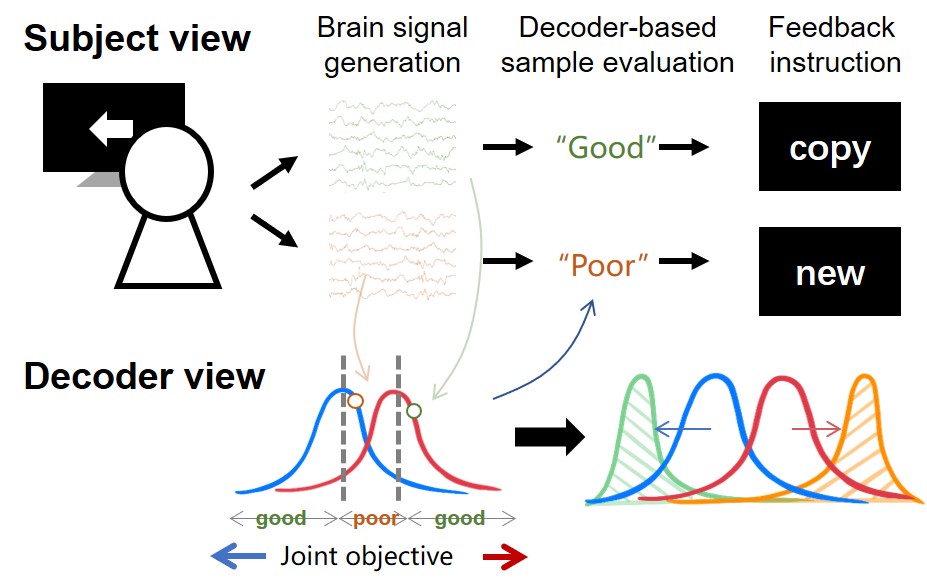} 
	\caption{The diagram of the human-machine joint learning process. From the subject's view, the subject tries to generate brain signals according to the instructions (such as left or right). At each trial, the system gives feedback (left or right) during the process, and evaluates the subject's brain signals according to the discriminative ability. For a trial with high signal quality, a ``copy'' command is given such that the subject should maintain the brain signal patterns; otherwise, a ``new'' command is given and the subject should change the way of thinking to improve the signal quality.}
	\label{joint}
\end{figure}

The optimization of the joint loss function relies on both the subject and the machine. On the one hand, the subject learns to maximize the discrimination ability between the two brain signal modes according to the decoder by optimizing $\theta_{H}$; on the other hand, the decoder tries to optimize a classification plane with $\theta_{M}$, where the brain signals from different modes can be maximumly classified. To solve the joint learning problem, we should re-model the learning for both humans and machines to cope with the jointly learning process. 

\subsubsection{Modeling of human learning \label{humanmodel}}

Here we model the learning process of the user. In a feedback learning process, the user adjusts the brain signals according to the feedback. In the human learning model, we simplify the problem by assuming that the subject generates new signals depending only on the last feedback. 
{\color{black} Then we can regard the process as a Markov process, which has been widely applied to modeling the human attention, decision-making or control system \cite{mugruza2022different,ghaderi2022neuro,busemeyer2020comparison,feinberg2012handbook}.}

Given the instruction $y$, the subject aims to generate brain signals in a certain mode. We divide the brain signals into two groups of ``good'' and ``bad/poor'' according to their discriminative ability with the decoder, as shown in Fig. \ref{joint}. We denote the ``good'' and ``bad/poor'' samples by $x_{G}$ and $x_{B}$. Thus, the objective of the subject is to increase the probability of generating $x_{G}$, and to decrease the probability of generating $x_{B}$.
The learning objective is therefore given by:
\begin{equation}
	\mathop{\arg\max_{\theta_{H}}} \frac{P(x_{G})}{P(x_{B})} ,
\end{equation}
where $P(x_{G})$ and $P(x_{B})$ represent the probability of good and bad samples of the subject. In the Markov sequential learning process, we can define a transition matrix of $x_{G}$ and $x_{B}$ by:
\begin{equation}
	\begin{bmatrix}
		P_{GG} & P_{GB} \\
		P_{BG} & P_{BB}
	\end{bmatrix}	=
	\begin{bmatrix}
		P_{GG} & 1-P_{GG} \\
		1-P_{BB} & P_{BB}
	\end{bmatrix}, 
\end{equation}
where $P_{ab}$ stands for the probability of transition from condition a to condition b. The steady-state vector could be derived as:
\begin{equation}
	\begin{bmatrix}
		P(x_{G})  & P(x_{B})
	\end{bmatrix}	=
	\begin{bmatrix}
		\dfrac{1-P_{BB}}{2-P_{GG}-P_{BB}} & \dfrac{1-P_{GG}}{2-P_{GG}-P_{BB}}
	\end{bmatrix}.
\end{equation}
Substituting the variables into the loss function, we get
\begin{equation}
	\mathop{\arg\max_{\theta_{H}}} \frac{P(x_{G})}{P(x_{B})} =	\mathop{\arg\max_{\theta_{H}}} \frac{1-P_{BB}}{1-P_{GG}}.
\end{equation}

According to the model, a high $P_{GG}$ and a low $P_{BB}$ are required for effective BCI learning.

\subsubsection{Modeling of machine learning}\label{PropAlgorithm}

Towards human-machine joint learning, we propose a novel classification algorithm based on a sample evaluation and adaptive learning of the decoder. 
The process of decoder learning has two folds. First, we evaluate the brain signals generated by the subject and determine whether a sample is ``good'' or ``bad'' for feedback in the sequential learning process. Second, the decoder should be dynamically tuned to cope with the new samples generated by the subject, where a novel training algorithm based on sample reweighting is proposed.
\begin{itemize}
	\item
	\textbf{Sample evaluation.}
	The determination of ``good'' and ``bad'' brain signals plays a key role in the process of human learning, and the evaluation of brain signal quality is the bond between the human and machine learning process. 
	
	Intuitively, ``good'' samples should contain high discriminative ability in the decoder's feature space. Thus, the determination of sample quality is given by the decoder, and the computation of sample quality can be diverse with different decoders. Here we take the SVM classifier as an example. With SVM, the determination of different classes is according to a sample's distance to the hyperplane constituted by the support vectors. In this manner, a sample far away from the hyperplane indicates low confusion in classification, thus it is considered a ``good'' sample, and samples with a small distance to the hyperplane are considered of low quality. Generally, the quality of a sample $i$ should be negatively correlated to its classification loss $L_i$. Therefore, we evaluate the quality of the sample $q_i$ as 
	\begin{equation}
		{q_i} \propto \frac{1}{L_i}.
		\label{weight}
	\end{equation}

	\item
	\textbf{Adaptive decoder learning with sample reweighting.}
	
	Given the quality of brain signal samples, the decoder assigns weights to the samples according to their quality to construct a better discriminative sample distribution, which serves as guidance for the subject to learn. 
	
	In typical machine learning approaches such as the SVM and Adaboost, samples with low quality (difficult samples) are more focused to improve the classification performance. While in the human-machine joint learning problem, we take the opposite strategy that focuses more on the ``good'' samples rather than the ``bad'' ones. 
	It is because the human-machine joint learning process forms a new problem that is quite different from typical ones. Instead of using fixed data, the sample generator (human) is intelligent and can adaptively cope with the decoder by learning to generate more ``good'' samples and fewer ``bad'' ones. By focusing on samples with high quality, the feedback of the decoder leads the subject to generate ``good'' samples to facilitate the learning process.
	Thus, the weight $v_i$ of the sample $i$ is assigned by
	\begin{equation}
		{v_i} = {q_i}.
		\label{weight2}
	\end{equation}

	Here we develop a novel adaptive learning approach by focusing on samples with high quality via a sample reweighting algorithm, which satisfies Eqs. \ref{weight} and \ref{weight2}. 
	Specifically, we design a novel loss function for the decoder to connect the loss and weight by the self-paced learning \cite{jiang2015self} algorithm, where the learning starts with an easy sample and gradually to difficult ones. The loss function of the decoder learning process can be described as:
	\begin{equation}
		\begin{aligned}
			\mathop{\min_{w,v}}&E(w,v;\lambda) = \\
			&\sum_{i=1}^{n} v_{i}L(x_i,y_i,w)+(1-\lambda)v_{i}-\frac{(1-\lambda)^{v_{i}}}{log(1-\lambda)},
			\label{loss3}
		\end{aligned}
	\end{equation}
	where $\sum_{i=1}^{n} v_{i}L(x_i,y_i,w)$ stands for the traditional loss function with weights $v$ and $$\sum_{i=1}^{n}(1-\lambda)v_{i}-\frac{(1-\lambda)^{v_{i}}}{log(1-\lambda)}$$ represents the penalty function $f(v;\lambda)$, which controls the pace of the self-paced learning. In this way, the weights are included in the loss function. 
	
	Since the loss could be taken as a biconvex function, an Alternative Convex Search \cite{kumar2010self} is used to optimize the loss function, which alternately finds the optimal solution to one group of variables by fixing the other variables. 
	\begin{itemize}
		\item
		\textbf{Optimization of $v$:}
		We first fix parameter $w$ and optimize $v$. The partial gradient of Eq. \ref{loss3} is:
		\begin{equation}
			\frac{\partial E(w,v;\lambda)}{\partial v_{i}}=L(y_{i},x_{i},w)+((1-\lambda)-(1-\lambda^{v_{i}})).
			\label{gradient}
		\end{equation}
		Let Eq. \ref{gradient} $=$ 0, and we can easily deduce:
		\begin{equation}
			log(L(y_{i},x_{i},w)+(1-\lambda))=v_{i}log(1-\lambda).
			\label{gradient2}
		\end{equation}
		So the solution for $E(w,v;\lambda)$ could be given by:
		\begin{equation}
			v'_{i}=\left\{
			\begin{aligned}
				& \frac{1}{log(1-\lambda)} log(L_{i}+(1-\lambda)),  \quad L_{i} < \lambda, \\
				& 0,  \quad {\rm otherwise},
			\end{aligned}
			\right.
			\label{solution}
		\end{equation}
		where $L_{i}$ stands for $L(y_{i},x_{i},w)$. As a result, we get the close-form solution for $v$. And in Eq. \ref{solution}, there is $v_{i}\propto \frac{1}{L_{i}}$, which means samples with higher loss will be given smaller weights, satisfying the requirement.
		
		\item
		\textbf{Optimization of $w$:}
		Given $v$, we fix $v$ and optimize $w$ as follows:
		\begin{equation}
			\frac{\partial E(w,v;\lambda)}{\partial w}=\frac{\partial \sum_{i=1}^{n} v_{i}L(x_i,y_i,w)}{\partial w}.
			\label{gradient3}
		\end{equation}
		As shown in Eq. \ref{gradient3}, the gradient is reduced to the partial derivative of a weighted loss problem. Close-form solutions can be derived if the loss function is hinge loss such as with the SVM. And for the loss of deep learning models, it also can be optimized through gradient descent.
	\end{itemize}
	
\end{itemize}

\subsubsection{Experimental paradigm for human-machine joint learning}\label{PropParadigm}

Given the modeling of human-machine joint learning, the problem is how to effectively feedback on the sample quality to the user. Here a ``copy/new" feedback strategy is proposed together with the paradigm of the joint learning process.

\begin{itemize}
	\item
	\textbf{The ``copy/new" feedback strategy.}
	We propose a novel ``copy/new'' feedback strategy to indicate whether the previous brain signal is ``good'' or ``poor''. In our paradigm, the trials are split into pairs and the instruction for the next trial depends on the quality of the previous trial. If the previous trial is of good quality, there would be a ``copy'' for the next trial. With the ``copy'' feedback, the subject is encouraged to generate brain signals similar to the previous one, which increases the $P_{GG}$. Otherwise, there would be a ``new'' instruction where the subject is asked to change the way of thinking and explore better signals, which helps decrease $P_{BB}$ in Section \ref{humanmodel}. {\color{black} In other words, the ``copy'' signals help subjects to maintain the previous ``good'' signal, and the ``new'' instruction will guide subjects to try other  possibilities.} The feedback lets the subject learn in a trial-and-error process, where the subject learns to shape the distribution of brain signals to a more discriminative condition with the guidance of the decoder. 
	\item
	\textbf{The paradigm of the joint learning process.}
	The diagram of the human-machine joint learning process is illustrated in Fig. \ref{joint} and Algorithm \ref{alg:jointlearning}. The subject tries to generate brain signals according to the instructions (such as left or right). Then the system gives the decoding results as feedback, meanwhile evaluating the quality of brain signals according to the discriminative ability. For a trial with high signal quality, a ``copy'' command is given such that the subject should maintain the brain signal patterns; otherwise, a ``new'' command is given and the subject should change the way of thinking to improve the signal quality. The aforementioned process constitutes a training block, and the process repeats several times in a training session. The decoder updates after each training session.
	
\end{itemize}

\begin{algorithm}[t]
	\caption{Pseudo code of the human-machine joint learning process}
	\label{alg:jointlearning}
	\textbf{Input}: Number of training sessions $n$, proposed algorithm $A$ in in Section \ref{PropAlgorithm} and proposed paradigm $P$ in in Section \ref{PropParadigm}\\
	\textbf{Output}: Fine-tuned classifier $C$ and trained subject $S$\\
	\begin{algorithmic}[1] 
		\STATE Let $k=1$.
		\STATE Calibration session of the paradigm without feedback.
		\STATE Update the classifier $C$ by $A$.
		\STATE $k=k+1$.
		\WHILE{$ k \leq n $}
		\STATE Online training with $P$.
		\STATE Update the classifier $C$ by $A$.
		\STATE $k=k+1$.
		\ENDWHILE
	\end{algorithmic}
\end{algorithm}

\begin{figure*}[!t]
	\centering
	\includegraphics[width=1.4\columnwidth]{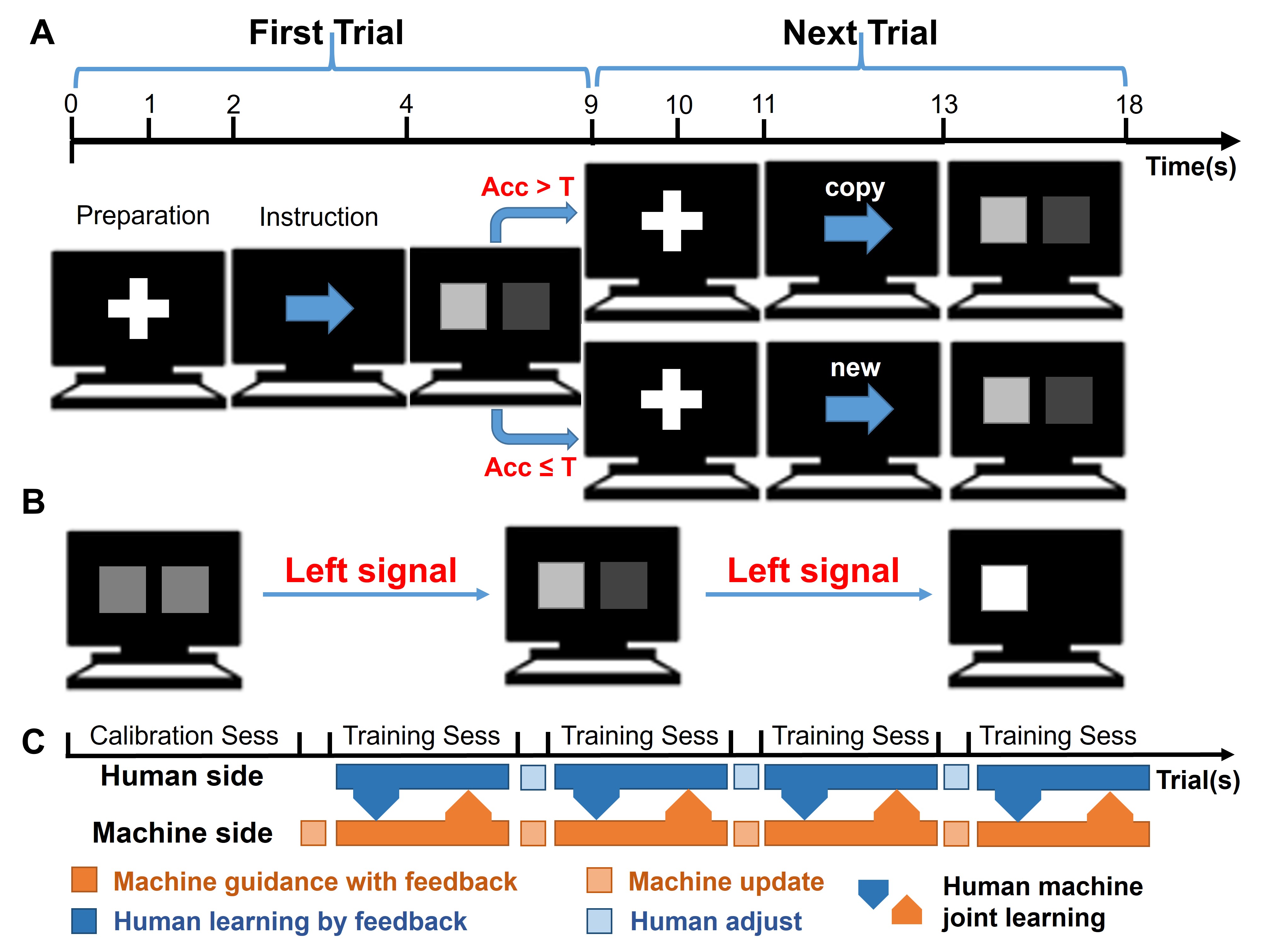} 
	\caption{The joint learning experiment paradigm with the MI task. (A) The trials are designed in a sequential manner as the First Trials and the Next Trials. For each trial, there are a 2-second-long preparation where a cross sign (+) is presented. Then an instruction arrow of ``left/right'' is presented for two seconds. With the Next trials, there are ``copy/new'' instructions along with the arrows in this stage. After that, the subject starts to perform MI as instructed for five seconds, during which feedback are given (as shown in B). (B) The feedbacks during the MI process include two squares indicating the decoding results for left and right respectively with the color of the squares. Lighter color indicates more discriminative classification performance. {\color{black} (C) The whole scheme of experiment including human side and machine side, which illustrates the joint learning process.}}
	\label{paradigm}
\end{figure*}

\subsection{The joint learning paradigm for motor imagery training} 

Here we specify the process of the joint learning paradigm with the MI task. 

\subsubsection{Experiment paradigm with the MI task \label{Experiment Paradigm}} 

The paradigm is presented in Fig. \ref{paradigm}. Subjects are asked to focus on the cross on the screen and perform left or right MI tasks as indicated by instruction arrows. The cross is shown on the screen for 2 seconds, turning from white to gray. After that, there is an arrow that points left or right, indicating the left or right MI task. The subject should perform MI for 5 seconds.
In the calibration session, the MI is performed without feedback. Subjects need to stay focused on the cross and keep their MI mental task without feedback. After the calibration session, the parameters of the algorithm are updated and feedback with ``copy/new'' instructions are provided. To reduce eye movement, we set the feedback as the brightness of the target. There are two squares spreading horizontally in the middle of the screen, in the left and right parts respectively. The subjects' task is to keep the indicated square as bright as possible in 5 seconds through MI. 
These feedbacks are denoted as $C^{n}_{left}$ and $C^{n}_{right}$, in which $n$ stands for the number of updates. $C^{0}_{left}$ and $C^{0}_{right}$ are set to 0.5 initially and updated as follows: 
\begin{equation}
	\begin{split}
		C^{n+1}_{left} &= C^{n}_{left} + \alpha((P_{left}-0.5)*2), \\
		C^{n+1}_{right} &= 1-C^{n+1}_{left},
	\end{split}
\end{equation}
where $P_{left} \in [0,1]$ is the posterior probability of left calculated by the online discriminator and $\alpha$ is a parameter to adjust the pace.

With the ``copy/new'' paradigm, the trials are split into pairs, namely the ``First trials" and the ``Next trials" (Fig. \ref{paradigm}). If the accuracy of the first trial achieves a threshold of $T$, there would be a ``copy'' along with the instruction arrow in the next trial, which informs subjects to keep the way of MI. Otherwise, these would be ``new'' in the screen above the instruction arrow, indicating that subjects should change the way of signal generation. The accuracy is calculated by the average accuracy of each slice during the online trial.

\subsubsection{MI decoding with sample reweighting}
{
	The decoding of MI brain signals consists of two stages. The first one is to extract effective features, where methods such as the common spatial patterns (CSP) and filter bank common spatial pattern (FBCSP) are used. And the second stage is to find a discriminator to classify the features. In order to deploy the proposed self-paced learning-based sample reweighting method, both stages should be taken into consideration.

	For the feature extraction, we used the weighted CSP by adding weights to the traditional CSP. Different from CSP, the normalized covariance $\overline{R_{1}}$ and $\overline{R_{2}}$ are calculated by weighted averaged over samples of each group:
	\begin{equation}
		R=\overline{R_{1}}+\overline{R_{2}}=\sum_{i} v_{1}^{i}r_{1}^{i} + \sum_{j} v_{2}^{j}r_{2}^{j},
		\label{normalizedcov}
	\end{equation}
	where $v_{1}^{i},v_{2}^{j}$ are the corresponding weight of samples in each group and
	\begin{equation}
		r_{1}^{i}=\frac{X_{1}^{i}X_{1}^{iT}}{trace(X_{1}^{i}X_{1}^{iT})}, \quad r_{2}^{j}=\frac{X_{2}^{j}X_{2}^{iT}}{trace(X_{2}^{j}X_{2}^{jT})},
		\label{covmatrix}
	\end{equation}
	in which $X_{1}^{i},X_{2}^{j} \in R^{N \times T}$ stand for the signals matrices of two conditions (left or right), $N$ denotes the number of channels, and $T$ is the number of samples per channel. 
	
	For the discriminator, we chose the SVM to cooperate with weights. Details of weighted-SVM could be found in \cite{lapin2014learning}, and the loss function is as follows:
	\begin{equation}
		L=\sum_{i=1}^{N}v_{i}(1-y_{i}f(x_{i})).
		\label{weightedhingeloss}
	\end{equation}
	Then the optimization problem becomes:
	\begin{equation}
		\begin{split}
			&\min_{\mathbf{w},b,\xi} \quad  \frac{1}{2}||\mathbf{w}||^{2}+\sum_{i=1}^{N}v_{i}\xi_{i} \\
			\mathrm{s.t.} \quad & y_{i}f(x_{i}) \geq 1-\xi_{i}, \quad \xi_{i} \geq 0 .
		\end{split}
		\label{weightedoptimization}
	\end{equation}
	This is the traditional weighted SVM problem, which could be solved easily.

	For the self-paced learning algorithm, we introduce new parameters $\Lambda$ and $\Delta \Lambda$ to control the learning pace more precisely, where $\Lambda$ stands for the proportion of samples we want to recruit for the training initially and $\Delta \Lambda$ is the incremental ratio of samples for training in every iteration. And the parameter  $\lambda$ is tuned by $\Lambda$ and $\Delta \Lambda$ in terms of the rank. The number of samples to be included in each iteration is specified by $\Lambda$ and $\Delta \Lambda$, and then $\lambda$ is calculated accordingly as \cite{jiang2014self}, which can be described as 
	the $1/\Lambda$ quantiles of loss $v$.
	
	Then we can design the workflow of our self-paced learning-based sample reweighting algorithm. The details of the algorithm are shown in Algorithm \ref{alg:algorithm} for MI decoding. Assuming we have a set of labeled data, we first randomly split the data into a training set $\left\{ X,Y \right\}$ of $N$ samples and a validation set $\left\{ X',Y' \right\}$.  At first, $\Lambda \times N$ samples are randomly selected from the training set. These data are used to train the initial weighted CSP and weighted SVM. After that, all data in the training set could get a weight, and the accuracy of the validation set is calculated. If  $ \Lambda < 1 $, which means not all the training set has been used, more data would be included in the update of weighted CSP and weighted SVM according to their weights. The iteration will continue until all the training set has been included. And the round with the highest accuracy on the validation set would be selected. The weight and parameters of weighted CSP and weighted SVM are the output.
	
	{\color{black} Also, the joint learning method could be employed in other machine learning methods, including deep learning. Under that condition, random selected $\lambda \times N$ samples will be fed to the neural network for initialization. After that, more and more samples would be assigned with different weights according to their loss to update the network by iteration. And the best model will be selected depending on the accuracy on $\left\{ X',Y' \right\}$, which is similar to the Algrithm \ref{alg:algorithm}.}

}

\begin{algorithm}[!h]
	\caption{The joint learning decoding algorithm for MI}
	\label{alg:algorithm}
	\textbf{Input}: Training set $\left\{ X,Y \right\}$ with $N$ samples, validation set $\left\{ X',Y' \right\}$. \\
	\textbf{Parameter}: $\Lambda$ and $\Delta \Lambda$, controlling the training pace \\ 
	\textbf{Output}: Weights: $v$, trained weighted-CSP and weighted-SVM\\
	\begin{algorithmic}[1] 
		\STATE Initialize $k=0$.
		\STATE Random select $\lambda \times N$ samples $\left\{ X_{k},Y_{k} \right\}$ from $\left\{ X,Y \right\}$.
		\STATE Train weighted-CSP on $\left\{ X_{k},Y_{k} \right\}$.
		\STATE Train weighted-SVM on $\left\{ X_{k},Y_{k} \right\}$.
		\STATE Update $v_{k}$ by Eq. \ref{solution}.
		\WHILE{$ \Lambda < 1 $}
		\STATE $ \Lambda=\Lambda+\Delta \Lambda$.
		\STATE $k=k+1$.
		\STATE $\lambda=1/\Lambda$ quantiles of $L$.
		\STATE Select data $\left\{ X_{k},Y_{k} \right\}$ by $ L_{i} < \lambda $.
		\STATE Update weighted-CSP on $\left\{ X_{k},Y_{k} \right\}$.
		\STATE Update weighted-SVM on $\left\{ X_{k},Y_{k} \right\}$.
		\STATE Test weighted-CSP and weighted-SVM on $\left\{ X',Y' \right\}$.
		\STATE Update $v_{k}$ by Eq. \ref{solution}.
		
		\ENDWHILE
		\STATE Select best iteration $K$ by accuracy on $\left\{ X',Y' \right\}$.
		\STATE \textbf{return} Weights at $K$ iteration: $v_{K}$, weighted-CSP and weighted-SVM at $K$ iteration
	\end{algorithmic}
\end{algorithm}

	\begin{figure*}[!t]
	\centering
	\includegraphics[width=2\columnwidth]{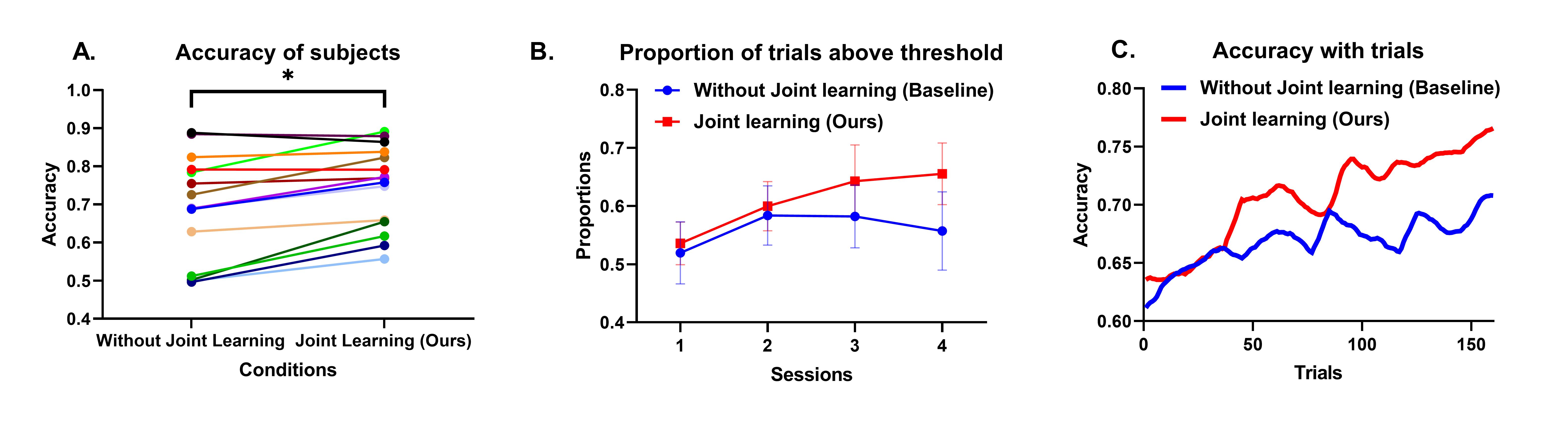} 
	\caption{Comparison of BCI control accuracy with and without joint learning. (A) The online accuracy of subjects with and without the joint learning process. $\ast$ indicates the significance with $t$-test (one-sided, $p \textless 0.005$). (B) The proportion of ``good'' trials across the sessions. (C) The trial-wise BCI control accuracy across trials.}
	\label{accchange}
\end{figure*}

\section{Experiments and results}
\subsection{Experimental settings}
We evaluated the proposed human-machine joint learning model with online BCI experiments in comparison with co-adaptive BCIs without joint learning.
{\color{black} A total of 18 subjects participated in the BCI experiments (eight females and ten males, age ranging from 23 to 27). 14 subjects took part in the experiment with CSP and SVM. And 4 subjects attended the experiment of EEGnet.} Each participant performed two experiments: a joint learning experiment and a co-adaptive learning experiment (baseline). To mitigate the effects of session order, the two experiments were arranged on two different days, at least two days apart. The order of the experiments was also randomly selected, where {\color{black} 9} subjects conducted joint learning first the rest performed co-adaptive learning experiments first. 
After each experiment, a simple questionnaire was filled out by the subject. The settings of the experiments were as follows:
\begin{itemize}
	\item \textbf{Joint learning experiment.} 
	There were a total of five sessions in the joint learning experiment. The first one was an initial calibration session without feedback, which had 60 trials including 30 left and 30 right trials in random order. After the initial calibration session, there were four joint learning sessions with feedback, consisting of 20 trials for each direction, and each trial was 9-second-long. The trials were presented in Fig. \ref{paradigm} and Section \ref{Experiment Paradigm}. For preprocessing, a four-order Butterworth Filter was adopted to filter the data from 8Hz to 30Hz. 
	The classifier was updated after every session. The threshold $T$ of ``copy/new'' was set to 70\%. $\Lambda$ was set to 0.2 for initialization and $\Delta \Lambda$ was set to 0.05 for updating the sample reweighting algorithm {\color{black} for SVM}. {\color{black} While for EEGnet, more samples were needed to initialize the training process, so $\Lambda$ was set to 0.5. And $\Delta \Lambda$ was set to 0.1 to reduce training iterations for saving time.}
	
	\item \textbf{Co-adaptive learning experiment without joint learning (baseline).} 
	We used the classical co-adaptive learning paradigm as the baseline for comparison. {\color{black}Specifically, the co-adaptive learning method in the paper is adopted from \cite{shenoy2006towards}, named as RETRAIN, which achieved the best performance. The same method was also used in \cite{abu2019co}.}
	The number of sessions in the co-adaptive learning experiment (control group) was the same as in the joint learning experiment. After the calibration session, the trials in the rest training session were provided with feedback from an {\color{black} SVM/EEGnet} decoder, which was also updated after every session.
\end{itemize}

For both the joint learning and the co-adaptive learning experiments, the software system was set up in Matlab, with the Psychtoolbox \cite{kleiner2007s}. {\color{black} And the computer was equipped with an Intel i7-10700k CPU, 64G RAM and an NVIDIA GeForce RTX 3080 GPU.} In CSP feature computing, three pairs of features were selected for classification. In the SVM classifier, a linear kernel was adopted. {\color{black} The time for inference was approximately 0.003 seconds. And the training of the algorithm cost 5.55$\pm$2.87 seconds. For EEGnet, we chose ADAM as the optimizer which is same to \cite{lawhern2018eegnet}. The inference time was approximately 0.004 seconds, while the training time was 153$\pm$23 seconds.} 

The feedback were calculated by the window of one second and updated every frame (1/60s). $\alpha$ was set to 0.2 for the feedback. For the offline analysis, we cut the five-second trial into four trial slices for one second, starting from 0.5 seconds and ending at 4.5 seconds after the instruction.

The EEG signals were recorded using a wireless EEG system (NSW24, Neuracle) with a sampling rate of 1000Hz. A total of 20 electrodes were placed on FC5, FC3, FC1, FCz, FC2, FC4, FC6, C5, C3, C1, Cz, C2, C4, C6, CP5, CP3, CP1, CPz, CP2, CP4 and CP6, according to the standard international 10-20 system. The ground and reference were FPz and CPz, respectively. The impedances of all electrodes were below the recommended value (10k$\Omega$) in the experiments.  

\subsection{Comparison of BCI control performance}

Here we compared the performance of the joint learning and the co-adaptive learning method both on SVM and EEGnet.  The performance was evaluated by the accuracy of online BCI control. We chose the highest accuracy among the last two sessions as the online BCI control accuracy, where the accuracy of sessions was averaged by the accuracy of all trials in the sessions.

\subsubsection{Analysis on SVM with joint learning}
Fig. \ref{accchange}A illustrated the online BCI control accuracy for both joint learning and co-adaptive learning with the fourteen subjects.  Overall, the BCI control accuracy of joint learning sessions was significantly higher than co-adaptive learning sessions without joint learning (paired $t$-test, $p \textless 0.005$). The average accuracies of methods with joint learning and without joint learning were 74.75\% and 69.06\%. For 11 out of 14 subjects, joint learning achieved a superior BCI control accuracy compared with the baseline co-adaptive experiments. The performance increases were most obvious with subjects with lower BCI control performance, indicating the joint learning process effectively facilitated the BCI training process.

In Fig. \ref{accchange}B, we examined the discrete performance by the proportion of successful trials. Specifically, we defined a trial with accuracy above the threshold $T$ (70\%) as a successful trial, which can be used as the discrete control signal.
Overall, the joint learning sessions obtained a higher proportion of successful trials compared with the baseline. 
With the joint learning system, the proportion of successful trials rose sustainingly over the four sessions, indicating the improvement of the subject's BCI control ability. Specifically, the proportion of successful trials increased to 60.0\%, 64.2\%, 65.5\% and  at the 2$^{nd}$, 3$^{rd}$, and 4$^{th}$ respectively.
While without joint learning, the proportion rose with the first two sessions, and the increase became trivial afterward. With session 4, the joint learning system outperformed the co-adaptive system significantly by 9.83\% (paired $t$-test, $p \textless 0.05$). 
\begin{figure*}[!t]
	\centering
	\includegraphics[width=2\columnwidth]{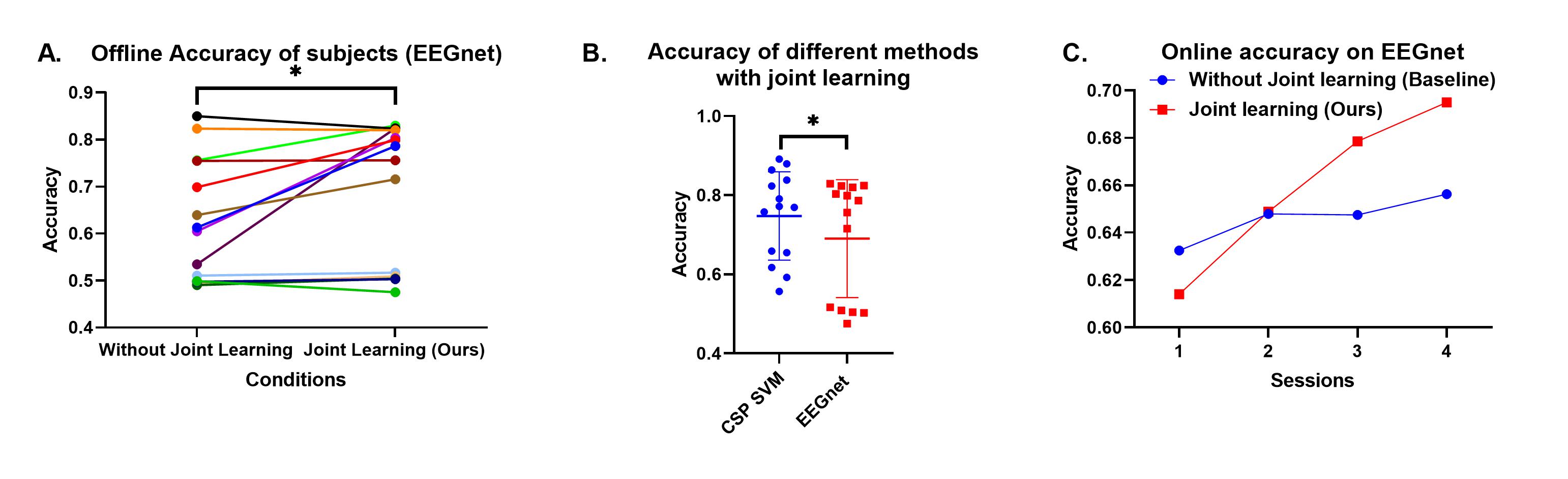} 
	\caption{\color{black}Comparison of BCI control accuracy with and without joint learning on EEGnet. (A) The offline accuracy of subjects with and without the joint learning process. $\ast$ indicates the significance with $t$-test (one-sided, $p \textless 0.005$). (B) Comparison between CSP SVM method and EEGnet. $\ast$ indicates the significance with $t$-test (one-sided, $p \textless 0.005$) (C) The online accuracy of EEGnet both with and without joint learning.}
	\label{EEgnet}
\end{figure*}
\begin{figure*}[!t]
	\centering
	\includegraphics[width=1.5\columnwidth]{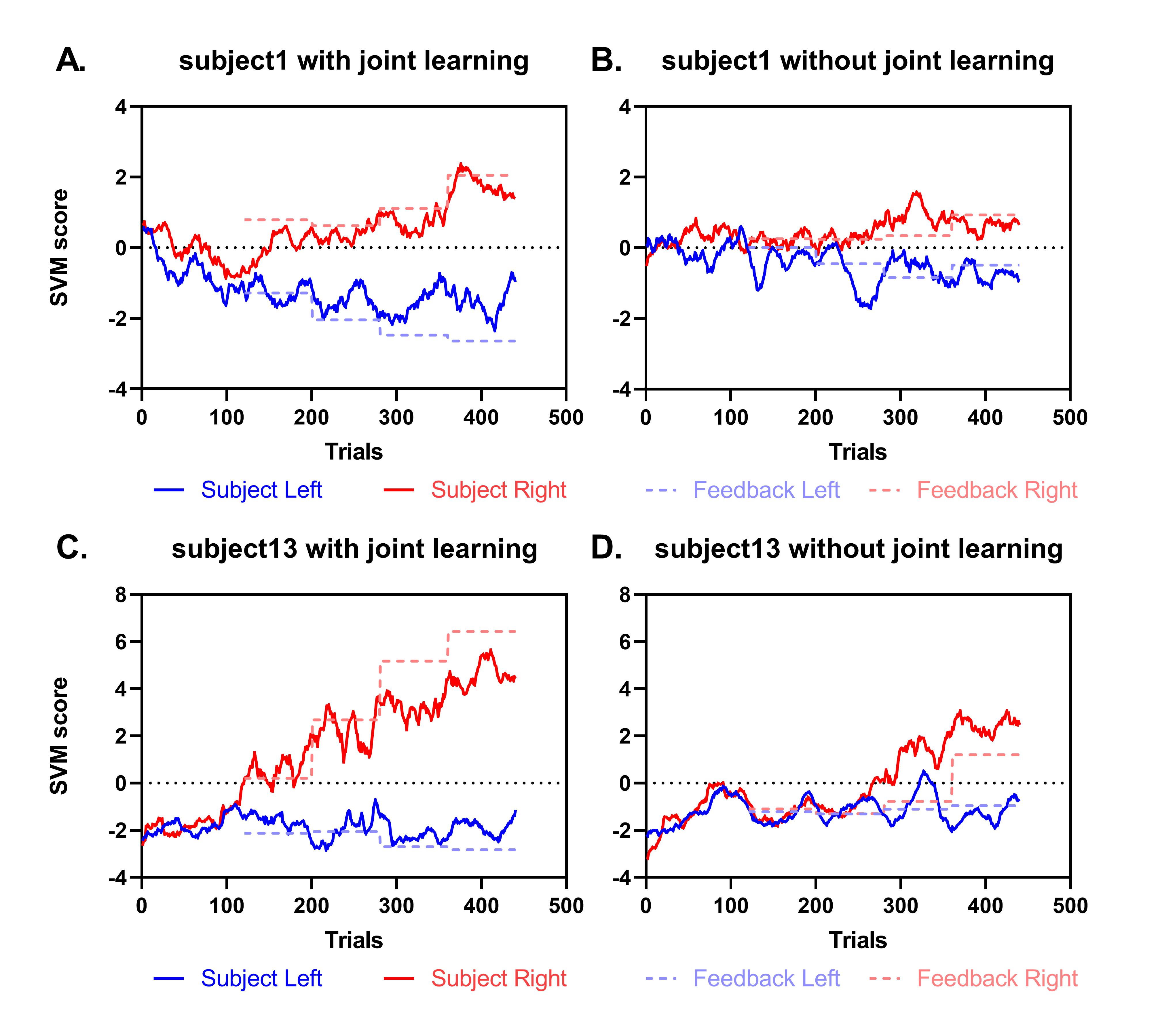} 
	\caption{\color{black}Analysis of the joint learning process. With the joint learning framework, subjects rapidly learn to generate well-separable signals (solid lines) with the guidance of the algorithm (dashed lines). Results on all subjects are in the Supplementary Material Fig.1.}
	\label{subject113camparisonsvmscore}
\end{figure*}
We further compared the trial-wise accuracy of BCI control. The trial-wise accuracy was computed by the mean accuracy of online accuracy during the trials, which was smoothed by the window of 10 points. Fig. \ref{accchange}C illustrates the average accuracy of every trial over all the subjects. For both joint and co-adaptive systems, the initial accuracies were around 60\%. With the learning process, the accuracy of joint learning rose rapidly to 76.58\%, while the accuracy of the baseline co-adaptive learning increased slower to 70.79\%. The results indicated that the subject learned more effectively with the joint learning system. 

{\subsubsection{Analysis on EEGnet with joint learning} \color{black}
We conducted both online and pseudo-online experiments on EEGnet with the proposed joint learning framework. The results are presented in Fig. \ref{EEgnet}. Specifically, we conducted pseudo-online experiments on the EEG data of 14 subjects, comparing the performance of the EEGnet algorithm with and without joint learning. As depicted in Fig. \ref{EEgnet}A, 11 out of 14 subjects demonstrated a significant increase in performance with the joint learning algorithm. Moreover, the average classification accuracy improved from 62.59\% to 69.03\% ($p \textless 0.005$), indicating that the proposed framework is effective and robust for both SVM and EEGnet algorithms. These findings suggested that joint learning framework help subjects to generate better signals and improved performance in MI-based BCI systems.

{\color{black}
Fig. \ref{EEgnet}B demonstrated the performance comparison between SVM and EEGnet methods. The results indicate a significant advantage of the CSP SVM method over the deep learning EEGnet method, with accuracies of 74.75\% and 69.03\%, respectively. This performance difference may be attributed to the fact that our framework starts the discriminator with a small number of samples, which can lead to overfitting in the neural network. }

{\color{black}
We further conducted online experiments on 4 subjects with the EEGnet and the results were shown in Fig. \ref{EEgnet}C. Both lines represent the online accuracy in the feedback sessions. Without the joint learning framework, the average accuracy increased slowly, from 63.24\% to 65.62\%. However, with the joint learning paradigm, subjects demonstrated a faster learning process, starting at 61.40\%, rising to 64.88\% and 67.85\%, ending at 69.49\%. These results provided evidence of the effectiveness of our proposed joint learning method and its flexibility in working with different algorithms.}}

\begin{figure*}[!t]
	\centering
	\includegraphics[width=2\columnwidth]{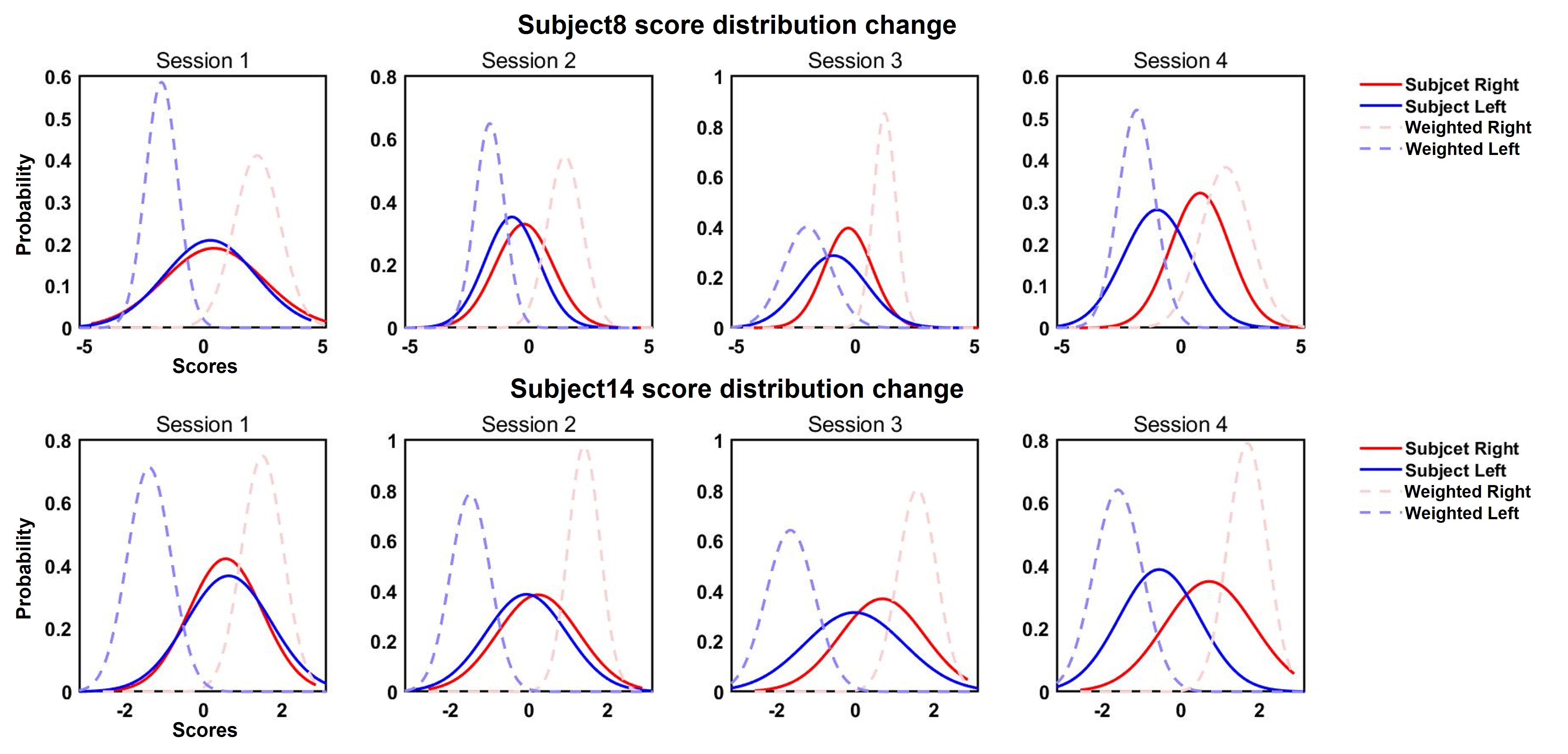} 
	\caption{\color{black} Changes of subject's brain signal distributions during the joint learning process. Results on all subjects are in the Supplementary Material Fig.2.}
	\label{distributionchange}
\end{figure*}

\subsection{Effectiveness of the joint learning method}
\subsubsection{Analysis of the joint learning process}
The major feature of human-machine joint learning is that the machine can guide the learning process of the subject for more efficient BCI training. In the MI task, the joint goal of the subject and the decoder is to maximize the discriminative ability of the ``left'' and ``right''  brain signal patterns on the decoder's space. 

Thus, we illustrated the discriminative ability of the ``left'' and ``right'' signals during the training process in Fig. \ref{subject113camparisonsvmscore}. The discriminative ability was evaluated by the SVM score, indicating the distance to the classification plane in the hyper-space. The blue and red lines stood for the scores of left and right respectively. With the joint learning process, the blue and red lines became distant rapidly, indicating the improvement of discriminative ability and the effectiveness of the joint learning process. Specifically, for subject1 in Fig. \ref{subject113camparisonsvmscore}A, the initial SVM scores for both ``left'' and ``right'' were around 0.5, which were difficult to discriminative by the decoder, indicating bad signal quality. In the training process, the ``left'' and ``right'' became more separated during training. After about 160 training trials (at the second training session), the SVM scores of brain signals for ``left'' and ``right'' were -1.59 and 0.42 respectively, which were well discriminative with the decoder. At the end of the training process, the brain signals for ``left'' and ``right'' were discriminative with SVM scores of -0.97 and 1.39 respectively. With the traditional co-adaptive learning process in Fig. \ref{subject113camparisonsvmscore}B, the learning process was inefficient compared with the joint learning process. Specifically, the brain signals were difficult to classify until the 250$^{th}$ trials, with SVM scores of -1.04 and 0.33. And the SVM scores were -0.88 and 0.65 at the end of training. With the subject 13, similar results were observed in Fig. \ref{subject113camparisonsvmscore}C and Fig. \ref{subject113camparisonsvmscore}D. It is worth noting that, for a fair comparison, the two chosen subjects had the experiments in a different order. Subject 1 did the baseline experiment first and subject 13 did the joint learning first.
The results strongly suggested that the joint learning process improved the efficiency of BCI learning effectively. 
\begin{figure*}[!t]
	\centering
	\includegraphics[width=2\columnwidth]{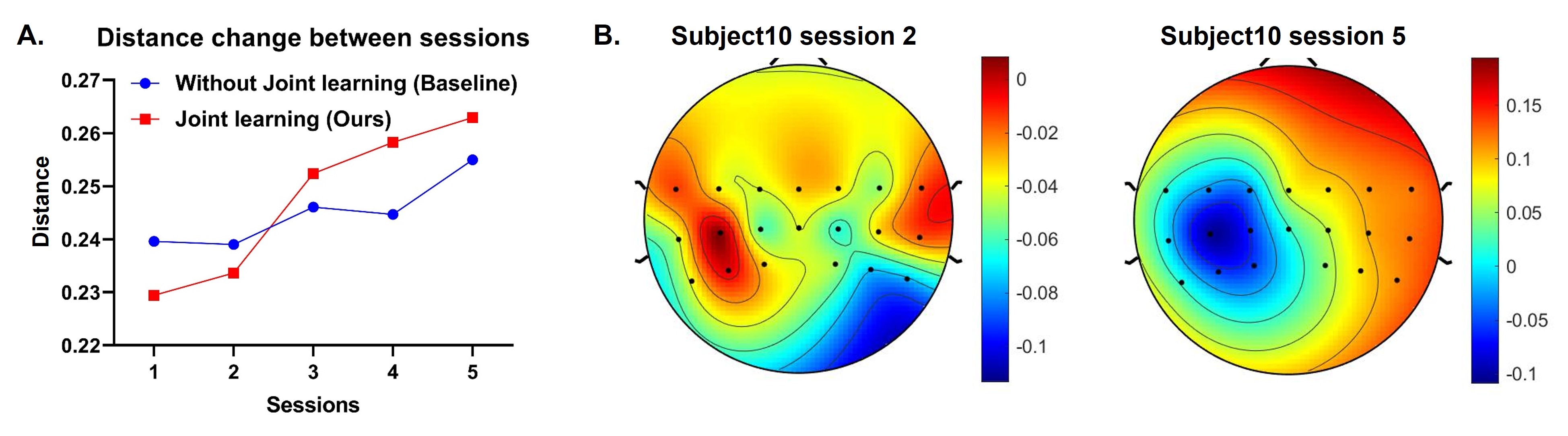} 
	\caption{\color{black} Signal changes of subjects. (A) Sample distance between the left and right signals across sessions. (B) Analysis of CSP features before and after joint learning. Results on all subjects are in the Supplementary Material Fig.3.}
	\label{signalchange}
\end{figure*}
\subsubsection{Analysis of the decoder's guidance to the subject}
Here we further investigated the effectiveness of the joint learning process by examining how the decoder guided the learning of the subject. In Fig. \ref{subject113camparisonsvmscore}, the dashed lines represented the mean of SVM scores accounting for the sample weights, which indicated the expected scores of subjects' signals by the decoder. In Fig.  \ref{subject113camparisonsvmscore}A and Fig.  \ref{subject113camparisonsvmscore}C, it was observed that the decoder's expectations by sample reweighting were with higher absolute values for both ``left'' and ``right'' brain signals, and the gap between the expected scores between the two classes became larger during the training process. It indicated that the brain signal quality improved with the training process, and the decoder adaptively tuned itself to cope with the learning of the subject. Also, the expectations from the decoder suggested the joint optimal point for both humans and machines. With the training process, the SVM scores of brain signals became closer to the optimal point, indicating the joint learning process drove both the subject and the decoder to learn toward the ``efficient'' direction effectively. While with the traditional co-adaptive approach, the decoder's expectations showed lower absolute values compared with the sample scores, thus the learning was mostly driven by the subject learning process, which degraded the efficiency of learning.

In order to illustrate the change of scores in detail, distributions of scores in different sessions across the experiments were shown in Fig. \ref{distributionchange}.  The red and blue solid curves stood for the distribution of left and right scores. And the dashed curves were weighted scores from feedback. All the distributions were fitted by Gaussian distributions.
Generally, the red and blue solid curves followed the separation of the dashed curves. With the dash curves moving more and more aside, the gap between solid curves also became more and more significant. This not only demonstrated the learning process of subjects but also presented the details of the guidance of the sample reweighting algorithm. Quantitatively, the differences of $\mu$ (peaks of the fitted distribution) for left and right distributions were computed to show the separation. For the subject1, the difference was 0.44 at the first session for the generated signal, but the difference for feedback was around 3.2. With the training process, the difference for generated signal gently rose to 2.1 in the last session. And the differences for feedback were always larger than it for generated signals, leading the way. The change was almost the same for subject13. The difference for generated signals grew from 0.1 to 4.7. While the difference in feedback increased from 5.1 to 6.6.
These changes meant subjects learned to generate better signals with more separable distributions with the guidance of feedback, explaining the process of joint learning.

\subsubsection{Analysis of brain signals with the learning process}

Here we analyzed how the brain signals of the subjects change during the learning process. We firstly illustrated the distance between ``left'' and ``right'' signals along with sessions in Fig \ref{signalchange}A. In this experiment, we re-calculated the CSP features offline with brain signals from the last two sessions, and the filters were applied to all 5 sessions for CSP feature computation. The distance was computed by: 
\begin{equation}
	distance=\dfrac{D_{inter}}{D_{intra}}
	\label{distance}
\end{equation}
where $D_{inter}$ stood for the inter-group distance, which was calculated by the Euclidean Distance between the mean points of two groups of CSP features. And $D_{intra}$ represented the intra-group distance, which was expressed by the Euclidean Norm of the standard deviation of two groups. We illustrated the average distance over 14 subjects in Fig. \ref{signalchange}A. 

With the training process, the distance between ``left'' and ``right'' brain features became more separable for both joint learning and the baseline approach, indicating the learning process of the subject. With the joint learning approach, the distance between ``left'' and ``right'' increased more rapidly from 0.229 to 0.263. The increase in joint learning was 0.034, which was 0.020 higher than the baseline approach without joint learning. In the last two sessions, the average distance between ``left'' and ``right'' brain features was 0.261 for the joint learning approach, which was 0.011 higher than the baseline method. Further, we illustrated the CSP features before and after the joint learning experiments of a subject in Fig. \ref{signalchange}B. For both before and after learning conditions, we selected the most discriminative CSP features for comparison. After the joint learning process, the subject learned the typical MI pattern with the CSP feature peaking around the C3 channel.

The results demonstrated that the joint learning approach could help the subject learn brain signal patterns both effectively and rapidly.

\begin{figure}[!t]
	\centering
	\includegraphics[width=1\columnwidth]{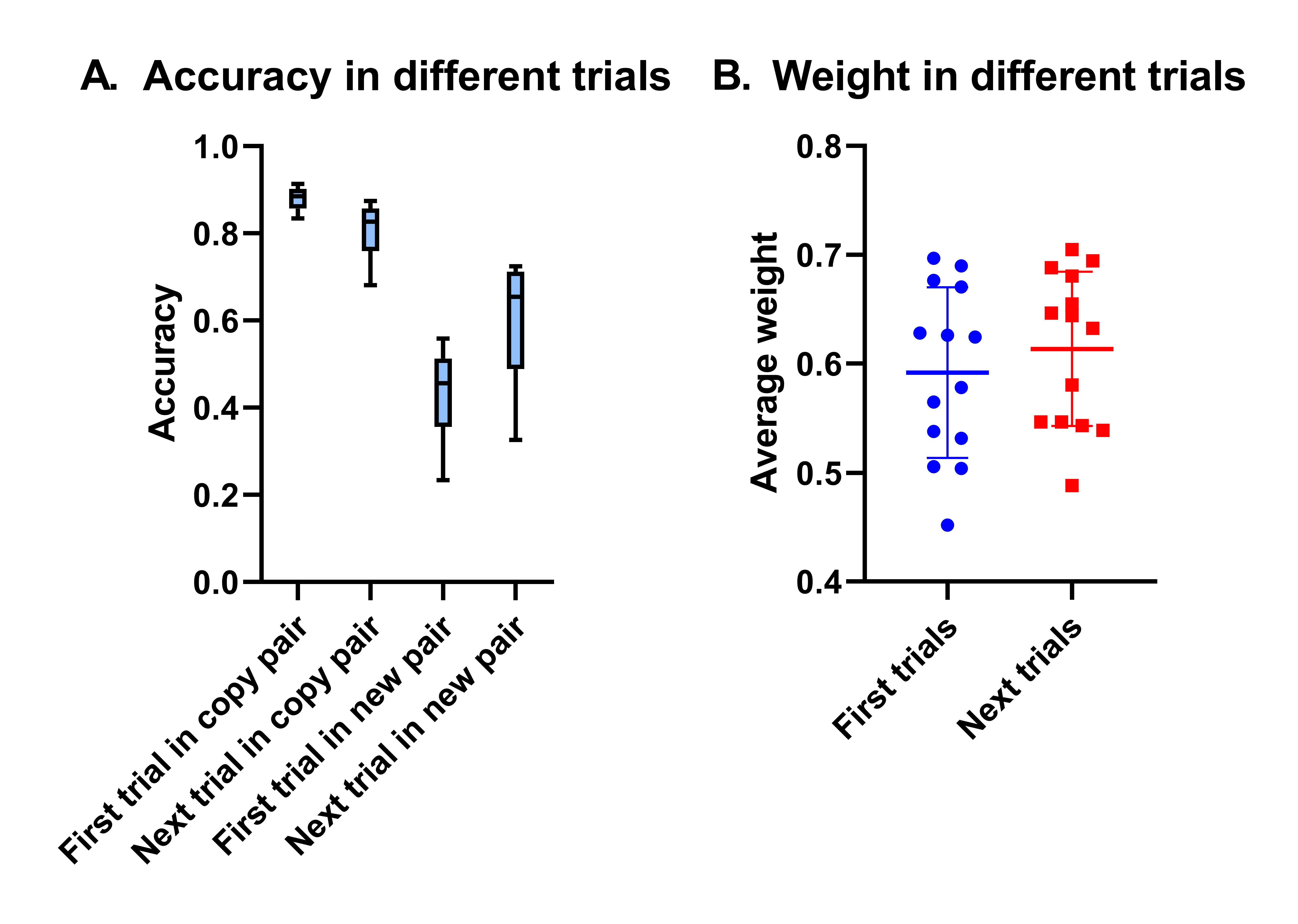} 
	\caption{Analysis of the ``copy/new" paradigm. (A) The accuracy of different trials. (B) Average sample weights of different trials.}
	\label{accracyindifferenttrials}
\end{figure}

\subsection{Effectiveness of the ``copy/new'' paradigm}
In this section, we evaluated the effectiveness of our ``copy/new'' paradigm from the aspects of accuracy and weight. We compared the averaged accuracy before and after a ``copy/new'' instruction. Specifically, we labeled the trials with four groups: 1) the first trial in a ``copy'' pair (before a ``copy'' instruction); 2) the next trial in a ``copy'' pair (after a ``copy'' instruction); 3) the first trial in a ``new'' pair (before a ``new'' instruction); 4) the next trial in a ``new'' pair (after a ``new'' instruction). 

Overall, the ``copy'' instruction guided the subjects to maintain a good brain signal pattern effectively, while the next trials after a  ``new'' instruction demonstrated a significant performance increase.
As shown in Fig. \ref{accracyindifferenttrials}A, the first trials in a ``copy'' pair demonstrated a high average accuracy of 88\%; after the ``copy'' instruction, the performance of the next trial slightly decreased while still keeping a high average accuracy of to 81\%. The results suggested that the ``copy'' instruction could guide the subject to repeat and keep a good brain signal pattern. With the ``new'' instruction, the first trials were with low accuracies of about 43\%. After the ``new'' instruction, the average accuracy of the next trial rose to 60\%. The results showed that the subjects tried to generate a new brain signal pattern different from the previous ones to seek a better performance, which was in line with expectations.

Furthermore, we analyzed the weights in different types of trials to show the cooperation between our paradigm and algorithm. Fig. \ref{accracyindifferenttrials}B presented the weight in different types of trials. The weights were calculated by averaging the weights on trials of the selected type. Overall, although the weights varied in a big range, the average weight on the next trials was slightly higher than the weight on the first trials. Specifically, the mean weight on the first trials was 0.58, and the mean weight on the next trials was 0.62 ($p \textless 0.05$ with one-sided t-test), which explained the better separability of the next trials. This indicated that the proposed ``copy/new'' instructions worked well with the sample reweighting algorithm, which also demonstrated the validity of the proposed learning paradigm.

\begin{figure*}[!t]
	\centering
	\includegraphics[width=2\columnwidth]{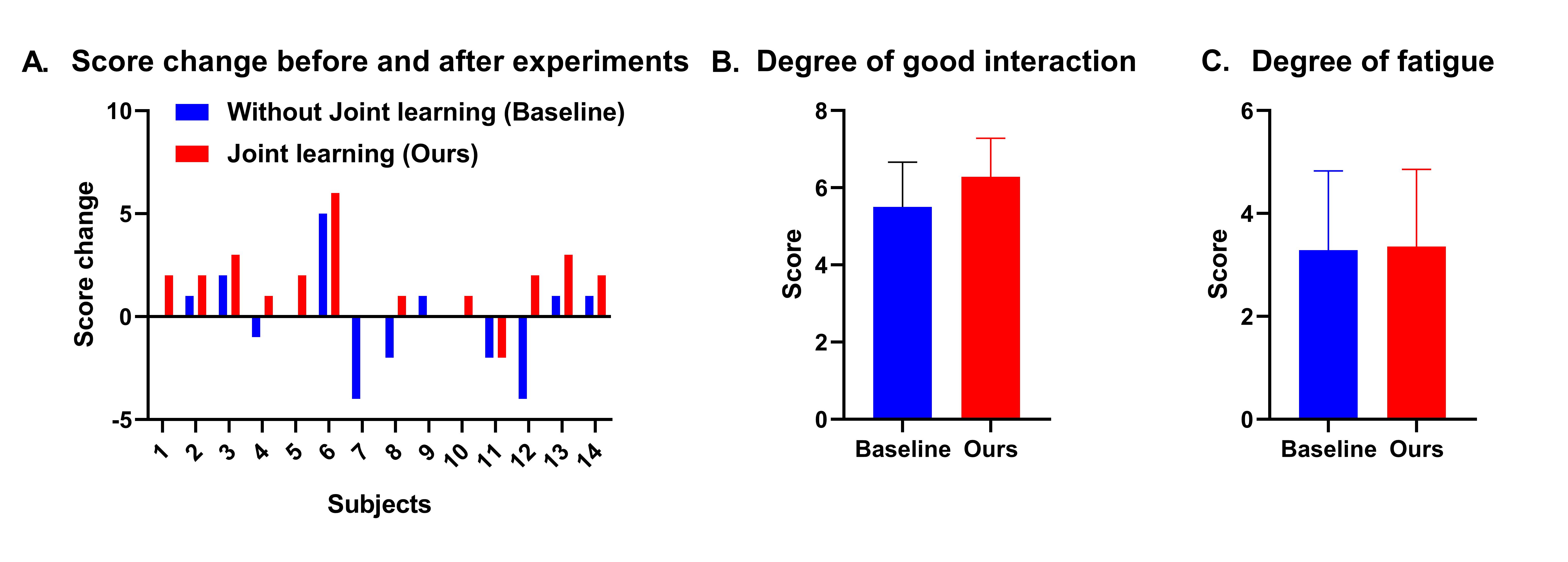} 
	\caption{Analysis of the questionnaires. (A) Scores before and after the experiments. (B) Scores of the interaction manner. (C) Scores of fatigue level.}
	\label{questionnaire}
\end{figure*}
\subsection{Analysis of the questionnaires}
The results of the questionnaire were presented to analyze from a subjective point of view.
The questionnaire consisted of six simple questions, measured by scores from one to seven, which were shown in the supplementary material in detail. The first four questions were about the ability of MI of the left and right hand respectively.  One pair was answered before the experiments and the other pair should be scored after the experiments (How do you think about your motor imagery ability of left/right hand before/after the experiment? one stands for very poor, and seven stands for very good). The result was plotted in Fig. \ref{questionnaire}A, where the score change was calculated by the sum of two questions after the experiments minus the sum before the experiments. It could be seen from the picture that red lines often had a higher value than blue lines, which meant subjects thought that they had better MI ability after our experiment with joint learning. In detail, the mean value of score change for the co-adaptive paradigm was -0.14, while for our joint learning paradigm, the mean change was 1.64. This meant subjects thought the co-adaptive training did not have significant effects, while our proposed joint learning method had a better training effect. This result was self-reported by subjects, which might be because the better control accuracy made subjects feel more confident.

Question 5 was about the degree of interaction, and level of knowing the feedback (Do you think you can understand the feedback? one stands for no, and seven stands for very good understanding). The results were in Fig. \ref{questionnaire}B. Our proposed method turned out to be slightly better than the control group, achieving an average score of around 6.29 (5.5 in the control group). This might be due to the ``copy/new'' instructions in our paradigm, giving subjects a stronger sense of participation. Also with the instructions, the subject can know about the separability or quality of generated signals more clearly. This also helped with the interaction during the experiments.

The last question was about fatigue (How much fatigue do you feel during the experiment? one stands for none, and seven stands for too much and intolerable). Fig. \ref{questionnaire}C showed the averaged score, in which scores of both experiments were almost the same (3.29 in the control group and 3.35 in our experiments). So, there was no significant difference between the efforts in the two experiments. Our experiments just helped subjects to change or hold their thoughts. This did not increase the workload of subjects.

\begin{figure*}[!t]
	\centering
	\includegraphics[width=2\columnwidth]{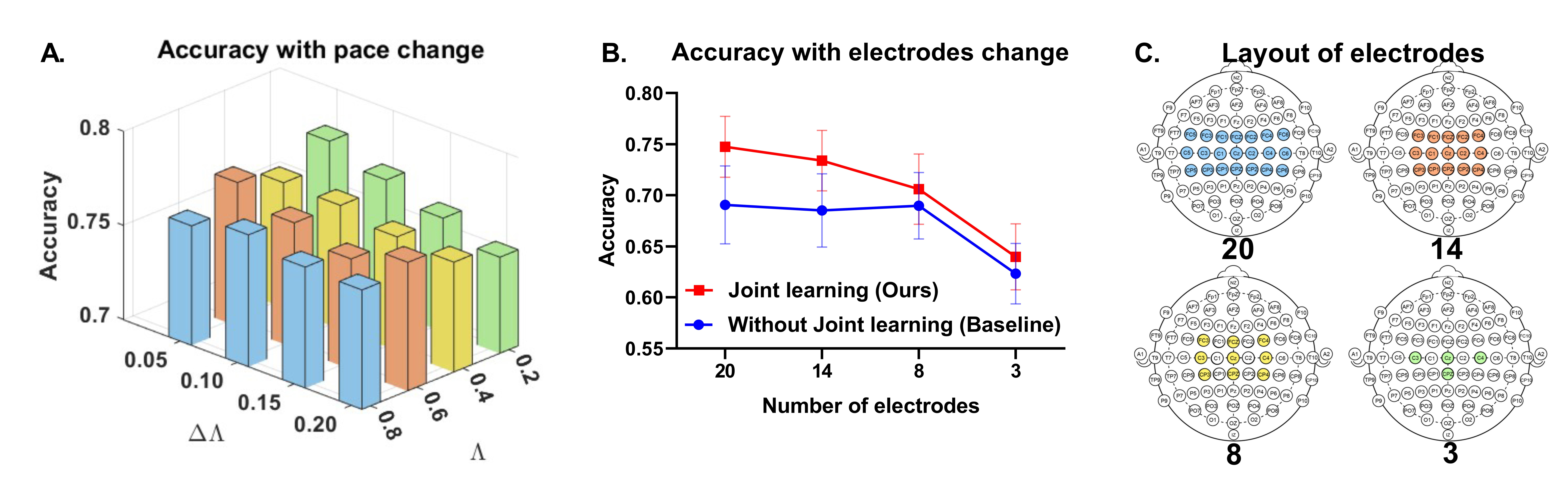} 
	\caption{\color{black} (A) Accuracy with different $\Lambda$ and $\Delta\Lambda$ (B) Accuracy with different number of electrodes. (C) Layout of electrodes.}
	\label{discussion}
\end{figure*}
\section{Discussion}
 {\color{black}\subsection{Novelty and contributions}

Spontaneous BCI training is often hindered by the training burden it poses. Existing co-adaptive learning methods, which adopt an alternate learning process, have shown suboptimal performance. In such methods, either the decoder or the user learns at a time \cite{shenoy2006towards, abu2019co}. However, human and computer algorithms are heterogeneous in information representation and learning mechanisms. Therefore, it is difficult to let humans and computers learn together in a uniform framework. In this study, we use brain signal samples generated by subjects as the intermedia, to enable the joint learning process between both sides. In the human learning process, the subject tries to generate “better” brain samples with the guidance of the computer, which tells the subject what are “good” samples. At the same time, the computer continuously updates itself adaptively to the human learning process. Thus, the learning process of both humans and computers can be optimized simultaneously. 
 
 In this paper, we first modeled the aforementioned joint learning process in a uniform model, with the Markov-based human learning process and the self-paced learning-based computer learning process. And then, we implement the framework by proposing a novel copy/new paradigm, to achieve the joint learning process. Online experiments demonstrated the effectiveness of the joint learning process.}
 
 
{\color{black}
\subsection{Sensitivity of parameters}
In this section, we conducted experiments to investigate the sensitivity of the algorithm's parameters. Specifically, we focused on two parameters, $\Lambda$ and $\Delta\Lambda$, which control the initialization and training pace of the algorithm. To evaluate their influence, we performed a grid search on the data from 14 subjects. We varied $\Lambda$ in the range of [0.2, 0.4, 0.6, 0.8] and $\Delta\Lambda$ in the range of [0.05, 0.10, 0.15, 0.20]. The results are presented in Fig. \ref{discussion}A. We observed that the accuracy ranged from 75.81\% to 78.11\%, with no significant difference found among the different parameter settings. The highest accuracy was achieved with $\Lambda=0.2$ and $\Delta\Lambda=0.05$, which is consistent with the parameter settings used in the online joint learning experiment. Conversely, the lowest accuracy was obtained with $\Lambda=0.2$ and $\Delta\Lambda=0.20$. Moreover, we found that the accuracy increased with the increment of $\Delta\Lambda$ when $\Lambda=0.2$. This suggests that when the number of initial samples is relatively small, more rounds are needed to reduce the effect of random initialization. In contrast, when more samples are included in the initialization, the results tend to be more stable.

\subsection{Electrode selection}

Electrode selection is a critical issue for motor imagery (MI) performance, in addition to parameters such as $\Lambda$ and $\Delta\Lambda$. The primary area for MI in the brain is the primary motor cortex (M1), which is located in the precentral gyrus of the frontal lobe \cite{pfurtscheller2001motor}. Therefore, electrodes of interest are typically C3 and C4, which are located over the left and right primary motor cortex, respectively \cite{neuper2009motor}. In the online experiments, we recorded data from 20 electrodes, along with CPz as a reference. As shown in Fig. \ref{discussion}C, the selected electrode set was centered at C3 and C4. To investigate the impact of electrode selection further, we reduced the number of electrodes from the outer side towards the center. We reduced the original 20 electrodes (FC5, FC3, FC1, FCz, FC2, FC4, FC6, C5, C3, C1, Cz, C2, C4, C6, CP5, CP3, CP1, CPz, CP2, CP4,CP6) to 14 electrodes (FC3, FC1, FCz, FC2, FC4, C3, C1, Cz, C2, C4, CP3, CP1, CP2, CP4), then 8 electrodes (FC3, FCz, FC4, C3, Cz, C4, CP3, CP4), and finally to 3 electrodes (C3, Cz, C4). The accuracy change is plotted in Fig. \ref{discussion}B. As expected, the accuracy decreased with the reduction of electrodes from 74.75\% to 63.97\%. The average accuracy of our proposed methods was always higher than the method without joint learning. However, the advantage diminished from 5.69\% to 1.64\%. This could be attributed to the decreasing number of electrodes resulting in less useful information. This result also helps to demonstrate the effectiveness of our proposed method.

\subsection{Limitations and future works}
\subsubsection{Multi-classification problem}
In this paper, we proposed novel a joint learning framework for effective BCI training, which took an MI system of the binary classification as an example. With the framework, different base algorithms could be deployed to deal with the training problem. For the multi-classification problem, our framework could easily fit to the problem. Because from the algorithm aspect, the key part is to evaluate samples according to their loss, to guide the human learning process. From the aspect of paradigm, the ``copy/new'' instructions would remain the same. As a result, the challenge lies in the base algorithm. In the experiments, we adopted the CSP SVM and the EEGnet as the base algorithms.
For the EEGnet, it was introduced to deal with all kinds of EEG problems. So it would be rather convenient to deal with multi-classification problem.
For the CSP SVM, it was designed for the binary classification problem \cite{ang2008filter}. However, there are some improvements to enhance CSP for multiclass MI \cite{kumar2020formulating}. 
With these new methods, our proposed framework promises to achieve good results on multi-classification problems.
\subsubsection{Markov process and endogenous attention}
As discussed in Section \ref{humanmodel}, the Markov process is a widely used framework for modeling human decision-making processes. Recently, it has been applied to the study of endogenous attention, as it can accurately reproduce reaction times in GO-GO experiments \cite{mugruza2022different}. Moreover, Markov models have been employed in the field of visual decision-making \cite{ghaderi2022neuro}, and hidden Markov models have been used to reveal endogenous neural activities that trigger perceptual changes by analyzing dynamic neural patterns \cite{lyu2022intrinsic}. In this paper, we model the learning process, including decision-making about MI strategies, using a Markov process. In future work, we aim to expand the model to include EEG patterns and features. This will provide further insights into human learning and decision-making processes in MI training.
}
{\color{black}\subsubsection{Applications on various EEG devices} 
The emergence of advanced sensor technology has facilitated the creation of compact and intelligent wearable electroencephalography (EEG) devices tailored for personal use \cite{gu2021eeg}. Devices with novel dry EEG sensors, which is convenient to wear and remove, have enhanced the usability of BCI systems \cite{liao2012biosensor}. Wireless devices with
dry and noncontact EEG electrode sensor have also been introduced \cite{chi2010wireless}. These devices are intended for real-life usage scenarios, which are significantly more complex than controlled laboratory environments. Consequently, ensuring the stability of BCI systems poses a considerable challenge. Our proposed human-machine joint learning framework exhibits vast prospects for application and further work in this field.}

\section{Conclusion}
Towards efficient and effective BCI training, the proposed human-machine joint learning framework enables the simultaneous learning of both the user and the decoder. We model the human-machine joint learning process in a uniform formulation and propose a novel joint learning framework for efficient BCI training. Online and psuedo-online experiments demonstrate the effectiveness of the proposed joint learning framework in rapid BCI learning. The proposed framework can be extended to MI tasks with more control degrees and also has the potential to guide the subject to generate new brain patterns out of the MI manner, to broaden the usability of BCI systems.

\section{ACKNOWLEDGMENT}
This work was partly supported by the grants from STI 2030 Major Projects (2021ZD0200400), and Natural Science Foundation of China (U1909202 and 61925603), and the Key Research and Development Program of Zhejiang Province in China (2020C03004).

\bibliographystyle{IEEEtran}
\bibliography{paper.bib}{}
\bibliographystyle{IEEEtran}
\end{document}